\documentclass[sigconf, table, xcdraw]{acmart} 

\AtBeginDocument{%
  \providecommand\BibTeX{{%
    \normalfont B\kern-0.5em{\scshape i\kern-0.25em b}\kern-0.8em\TeX}}}

\setcopyright{none}
\acmYear{}
\acmDOI{}

\acmConference[CARRV 2021]{CARRV 2021: Fifth Workshop on Computer Architecture Research with RISC-V}{June 17th, 2021}{Worldwide}
\acmBooktitle{CARRV 2021: Fifth Workshop on Computer Architecture Research with RISC-V,
  June 17th, 2021, Worldwide}
\acmISBN{}

\usepackage{csvsimple}
\usepackage{booktabs}
\usepackage{enumitem}
\usepackage{pbox}
\usepackage{acronym}

\usepackage{tikz}

\usepackage{amsmath}
\usepackage{amsfonts}
\usepackage{mathtools}
\usepackage{listings}

\definecolor{codegreen}{rgb}{0,0.6,0}
\definecolor{codegray}{rgb}{0.5,0.5,0.5}
\definecolor{codepurple}{rgb}{0.58,0,0.82}
\definecolor{backcolour}{rgb}{0.95,0.95,0.92}

\lstdefinestyle{mystyle}{
    commentstyle=\color{codegreen},
    keywordstyle=\color{magenta},
    numberstyle=\tiny\color{codegray},
    stringstyle=\color{codepurple},
    basicstyle=\ttfamily\footnotesize,
    breakatwhitespace=false,         
    breaklines=false,                 
    captionpos=b,                    
    keepspaces=true,                 
    numbers=none,                    
    numbersep=5pt,                  
    showspaces=false,
    showstringspaces=false,
    showtabs=false,                  
    tabsize=2,
    lineskip=0pt,
    frame=single
}

\makeatletter
\newcommand{\verbatimfont}[1]{\renewcommand{\verbatim@font}{\ttfamily#1}}
\makeatother

\newcommand{\HPMcounter}[1]{{\texttt{\fontsize{8pt}{9pt}\selectfont #1}}}

\begin{document}
\title{Supporting RISC-V Performance Counters through Performance analysis tools for Linux (Perf) }

\author{Joao Mario Domingos}
\orcid{0000-0002-3038-1073}
\affiliation{INESC-ID, Instituto Superior Técnico - Universidade de Lisboa, Portugal}
\email{joao.mario@tecnico.ulisboa.pt}

\author{Pedro Tomas}
\orcid{0000-0001-8083-4432}
\affiliation{INESC-ID, Instituto Superior Técnico - Universidade de Lisboa, Portugal}
\email{pedro.tomas@inesc-id.pt}

\author{Leonel Sousa}
\orcid{0000-0002-8066-221X}
\affiliation{INESC-ID, Instituto Superior Técnico - Universidade de Lisboa, Portugal}
\email{las@inesc-id.pt}


\renewcommand{\shortauthors}{Domingos, J.M.; Tomas, P.; Sousa, L.}


\begin{abstract}
    Increased attention to RISC-V in Cloud, Data Center, Automotive and Networking applications, has been fueling the move of RISC-V to the high-performance computing scenario. However, lack of powerful performance monitoring tools will result in poorly optimized applications and, consequently, a limited computing performance. While the RISC-V ISA already defines a hardware performance monitor (HPM), current software gives limited support for monitoring performance. In this paper we introduce extensions and modifications to the Performance analysis tools for Linux (perf/perf\_events), Linux kernel, and OpenSBI, aiming to achieve full support for the RISC-V performance monitoring specification. Preliminary testing and evaluation was carried out in Linux 5.7 running on a FPGA-booted CVA6 CPU, formerly named Ariane, showing a monitoring overhead of 0.283\%.
\end{abstract}

%
%



\maketitle


    \newacro{ALU}[ALU]{Arithmetic and Logic Unit}
    \newacro{API}[API]{Application Programming Interface}
    \newacro{ASC}[ASC]{Texas Instruments Advanced Scientific Computer}

    \newacro{CISC}[CISC]{Complex Instruction Set Computer}
    \newacro{CPU}[CPU]{Central Processing Unit}
    \newacro{CSR}[CSR]{Control Status Register}
    \newacro{CTI}[CTI]{Cycle, Time and Retired Instructions}

    \newacro{DMA}[DMA]{Direct Memory Access}
    \newacro{DSL}[DSL]{Domain Specific Language}
    \newacro{DLP}[DLP]{Data-Level Parallelism}
    \newacro{DAE}[DAE]{Decoupled Access-Execute}
    
    \newacro{FP}[FP]{Floating-Point}
    \newacro{FPGA}[FPGA]{Field Programmable Gate Array}
    
    \newacro{GPU}[GPU]{Graphics Processing Unit}
    \newacro{GPGPU}[GPGPU]{General Purpose Graphics Processing Unit}

    \newacro{HPM}[HPM]{Hardware Performance Monitor}
    \newacro{HPC}[HPC]{High-Performance Computing}

    \newacro{ISA}[ISA]{Instruction Set Architecture}
    \newacro{IS}[IS]{Instruction Set}
    \newacro{IST}[IST]{Instituto Superior T\'ecnico}
    \newacro{IoT}[IoT]{Internet of Things}
    \newacro{ILP}[ILP]{Instruction-Level Parallelism}
    \newacro{IPC}[IPC]{Instructions per Clock}

    \newacro{PMU}[PMU]{Performance Monitoring Unit}

    \newacro{RISC}[RISC]{Reduced Instruction Set Computer}
    
    \newacro{SIMD}[SIMD]{Single Instruction Multiple Data}
    \newacro{SISD}[SISD]{Single Instruction Single Data}
    \newacro{SIMT}[SIMT]{Single Instruction Multiple Thread}
    \newacro{SP}[SP]{Single-Precision}
    \newacro{SVE}[SVE]{Scalable Vector Extension}
    \newacro{SCROB}[SCROB]{Stream Configuration Reorder Buffer}
    
    \newacro{RTL}[RTL]{Register-transfer level}
    \newacro{RVV}[RVV]{RISC-V "V" Vector Extension}
    \newacro{TLP}[TLP]{Task-Level Parallelism}

    \newacro{SAXPY}[SAXPY]{Single-precision computation of the product of A with each element of matrix X added to the respective element of the matrix Y. - simple 1-D memory access, low in compute intensity.}
    \newacro{MEMCPY}[MEMCPY]{Memory copy. Procedure that copies a source zone of memory to a destination one.}
    \newacro{STAR100}[STAR-100]{Control Data Corporation STrings of binary digits that made up ARrays - 100}
    
    \newacro{UVE}[UVE]{Unlimited Vector Extension}

    \newacro{VL}[VL]{Vector-Length}
    
    \newacro{IRSmk}[IRSmk]{Scientific microkernel, used in radiation simulations - very intensive 3-D memory accesses, with low computational effort.}
    \newacro{HACCmk}[HACCmk]{Scientific microkernel, used in universe simulation - simple memory accesses with intensive computation.}
    
    \newacro{FIFO}[FIFO]{First-In First-Out}
    \newacro{SW}[SW]{software}
    \newacro{HW}[HW]{hardware}    


\section{Introduction} \label{sec:Introduction}

In a High-Performance Computing era, naively porting workloads across computing platforms results in limited computing performance. Naturally, many factors come into play when justifying the observed performance limitations, including poor software implementations that result in high computational complexities, in inefficient data structures and/or limited cache usage, resulting in a memory bound execution, in execution bottlenecks, leading to processor stalls, etc.


Hence, to optimize an application, one must first monitor the its execution, identify the main performance bottlenecks, and then tailor the software to best fit with the underlying hardware. Naturally, this cannot be performed using sheer performance metrics (e.g., execution time or clock cycles), as multiple factors come into play when mapping the software to a modern computing system (e.g., in- vs out-of-order execution engines, pipeline stages, execution ports and corresponding latencies, re-order buffers, load/store queues, cache organization, etc). Consequently, we easily observe that, in the current computing scenario, there is an increasing need to capture and analyse detailed performance metrics, in order to allow in-depth architecture modeling and optimization procedures (e.g. \cite{marques2020application}).

While Intel and ARM have proprietary performance monitoring solutions \cite{kleen_intel_2015,su_multi-core_2011,zeinolabedin_real-time_2021,lee_using_2017}, which allow software developers to take the ultimate advantage of their hardware, RISC-V is still dependent on custom/vendor-specific solutions with no complete support for common performance monitoring software tools, such as PAPI \cite{jagode_papi_2019,dongarra_using_2001,london_papi_2001} or Oprofile \cite{oprofile_oprofile_2021}. Mainly, RISC-V is still not fully supported in the Linux kernel monitoring tool Perf \cite{weaver_linux_2013,de_melo_new_2010,kao_supporting_2018}. Only fixed counters are currently supported without event configuration, and no control over the counters is provided (e.g., pausing, enabling, disabling).

To improve the performance analysis tools for Linux (Perf) with support for the RISC-V performance monitoring facilities, in this paper we propose the following software additions and modifications:
\begin{itemize}
    \item Support the latest RISC-V HPM specification in the Linux Perf Kernel Driver;
    \item Introduce RISC-V events, configurable through Perf;
    \item Support multiple platforms with distinct sets of events;
    \item An OpenSBI extension for privileged interaction with the RISC-V performance monitoring hardware.
\end{itemize}

Considering the multiple available RISC-V implementations, and their dissociated performance monitoring hardware implementations, we propose to consider coping strategies such as backwards compatibility and implementation features discovery. Even so, it is not possible to encompass all the details and specializations of the available implementations and RISC-V specifications at once, and therefore, we set specification version 1.11 \cite{watermanRISCVInstructionSet2019a} as our primarily supported target, and try to make the software flexible to support the majority of implementations.

This document represents early work that could be subjected to changes due to future knowledge and realizations. In a final form of the work, our aim is to arrive at a complete support for Perf and eventually for other relevant performance monitoring and optimization software (e.g., PAPI).

\section{RISC-V Performance Monitor} \label{sec:RVHPM}
\begin{figure*}[t]
    \centering
    \includegraphics[width=0.8\textwidth]{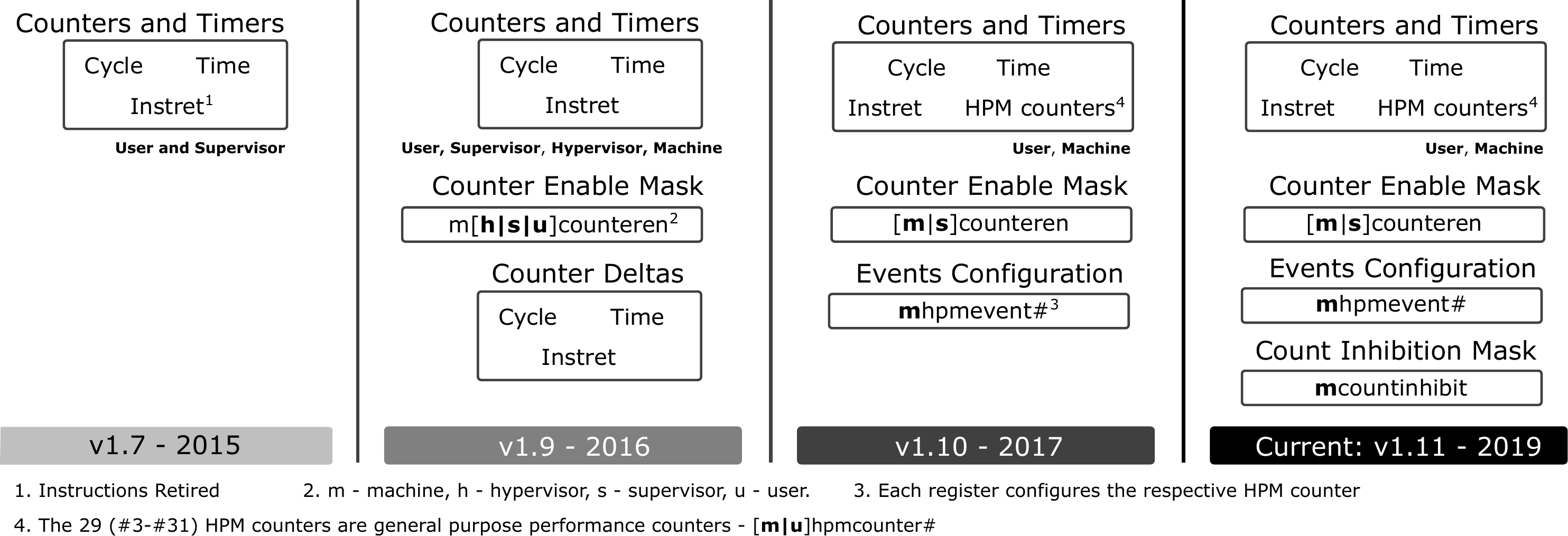}
    \caption{Evolution of the RISC-V Hardware Performance Monitor specification}
    \label{fig:RISC-V_HPM_Overview}
\end{figure*}

The RISC-V ISA has seen a continuous ecosystem development, from wider software compatibility to an increasing number of hardware implementations \cite{lupori2018towards,celio_boom_2017,zhao_sonicboom_2020,balkind_openpitonariane_2019,matthews_taiga_2017,zaruba_cost_2019,sifive_sifive_2018,chen_xuantie-910_2020}. Alongside the software and hardware, the RISC-V specification also shows an uninterrupted evolution, driven by the growing requirements of the RISC-V ecosystem. Since RISC-V privileged specification version 1.7 \cite{waterman_risc-v_2015}, a minimal performance monitoring interface was defined (see \autoref{fig:RISC-V_HPM_Overview}). From then, the specification has introduced additional counters and necessary features for access control and event multiplexing.

\subsection{Early specifications}
The first RISC-V privileged specification, version 1.7, introduced the first attempt at monitoring core's performance. The implementation, supporting three fixed counters \ac{CTI}, allowed for baseline performance monitoring of a RISC-V implementation, enough for calculating the \ac{IPC} metric. At v1.7, the \ac{PMU} had all the counters accessible at user and supervisor privileges, lacking control over non-privileged access.

Version 1.9 \cite{waterman_risc-v_2016} introduced control over the privileged counter accesses, a counter-enable mask was introduced by means of three registers accessible only at machine-level, and imposing read control over the \ac{CTI} counters at hypervisor, supervisor and user level. In addition, v1.9 introduced a set of counter deltas, a counter would keep the difference between each of the lower privilege counters and the respective machine-level counter (e.g., stime-mtime=mstime\_delta). These delta counters were removed after version 1.9. At the time, RISC-V performance monitoring was still limited to the set of three fixed registers, without support for general purpose or fixed-event performance monitoring registers.

\subsection{Configurable events and counters}
Support for 29 additional performance monitor registers was introduced with version 1.10. The \ac{HPM} counters, ranging from \HPMcounter{hpmcounter3} to \HPMcounter{hpmcounter31}, can be individually configured by setting an event identifier in the corresponding \HPMcounter{hpmevent} registers, a set of XLEN-bits registers (e.g., XLEN = 64 in a 64-bits implementation). This amounts for, virtually, $2^{64}$ selectable events for a single register, a value that surpasses any realistic implementation, providing an overly large design flexibility. The RISC-V specification states that the number, width and supported events of each \HPMcounter{hpmcounter} is platform-/implementation-specific. Even so, HPM counters are limited to a maximum width of 64-bits.

When setting the \HPMcounter{hpmevent} registers, event 0 is considered as the null event, and both the event configuration and the counter registers can be hardwired to 0, indicating that no event counting can occur.
Each event counter (\HPMcounter{hpmcounter\#}) is writable in an WARL (write any, read logical) scheme, allowing for each counter to be individually reset/set \cite{waterman_risc-v_2016-1}.

\subsection{Additional Features and Future Objectives}

In the latest ratified specification, version 1.11 \cite{watermanRISCVInstructionSet2019a}, individual counter inhibition (i.e., stop counting) was introduced, allowing the software to atomically sample events. This is accomplished through the introduction  of the \HPMcounter{mcountinhibit} register, where each of the 32-bits can be set to inhibit the respective \ac{HPM} counter.

Current specifications suggest that future versions could include support for common event standardization, as to count ISA-level metrics, such as executed floating-point or integer instructions. Similarly, some very common and widely supported micro-architectural metrics could be standardized (e.g, L1 instruction cache misses). Another feature that may appear in future specifications is the support for counter overflow interrupts, allowing the software to accurately count events that overflow the counters at a faster pace than the event sampling occurs. Although, the occurrence of such continuous overflowing is unlikely, considering implementations with 64-bits counters.

\subsection{Summary}

At the time, the RISC-V \ac{HPM} is still significantly less complex than the x86 counterpart \cite{intel_intel_2016} and not comparable to the dedicated performance analysis tools like ARM's coresight and Intel's PCM-based monitoring solutions \cite{kleen_intel_2015,su_multi-core_2011,zeinolabedin_real-time_2021,lee_using_2017}. Even so, the RISC-V HPM specification is a flexible generic performance monitoring solution, and being open-source allows any degree of implementation freedom. Considering the current state of the RISC-V privileged specification, we propose, in the following section, an approach to monitor the performance counters in RISC-V through Linux Perf.
\section{Proposed Approach and new extensions} \label{sec:Proposed}

    
    
    
    

This work starts from the current Linux Perf implementation, developed after the RISC-V privileged specification version 1.10. This Perf implementation provides basic support for adding, deleting, starting and stopping software-side events. However, a significant limitation is the inability to write to counters and event configuration registers, as those writes require machine-level privilege, not available without a dedicated OpenSBI extension. Due to the writing limitation, it is not possible to configure events in a specific counter, significantly limiting Perf to the fixed set of \ac{CTI} counters \cite{kao_supporting_2018}.

\begin{figure*}[t]
    \centering
    \includegraphics[width=0.86\textwidth]{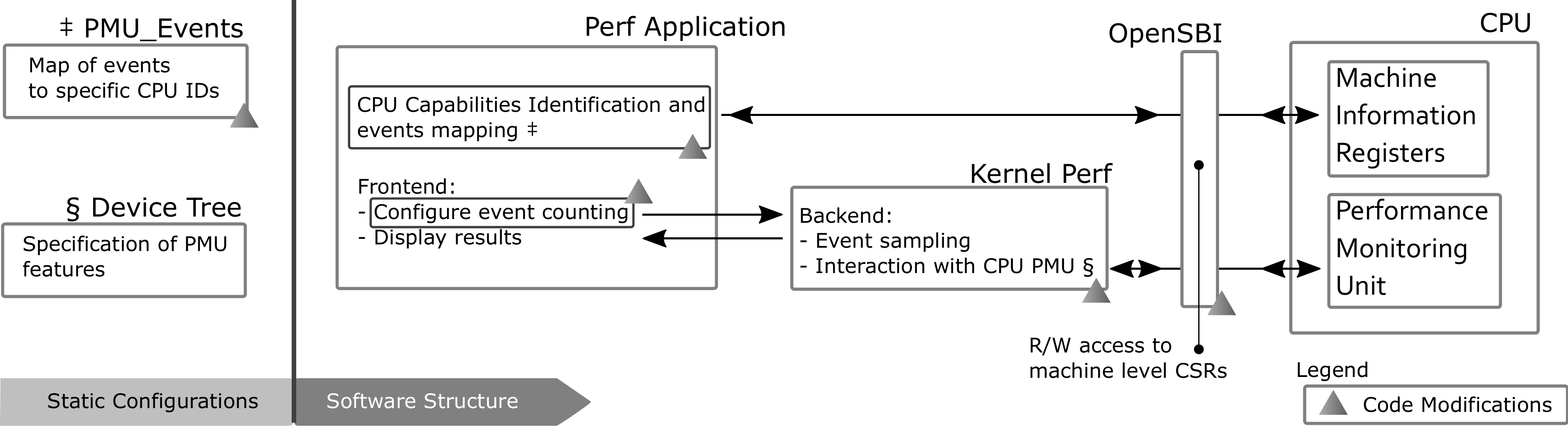}
    \vspace*{-2mm}
    \caption{Overview of the system software structure}
    \label{fig:SystemOverview}
\end{figure*}

Considering the current limitations, our proposal starts by providing a mechanism to write and read on machine-level privileged counters and registers, through the introduction of a new OpenSBI extension. Additionally the kernel Perf driver and the Perf tool were also modified. The Linux performance monitoring system is divided into the Perf application and the kernel driver, both connected through the \texttt{perf\_event\_open} system call, where the kernel driver samples the events from the performance monitoring hardware counters. An overview of the system, alongside with proposed modifications, is depicted in \autoref{fig:SystemOverview}.

\subsection{OpenSBI HPM Extension}
To define an interface between software and hardware and provide the required privileged access to machine-level registers, the Hardware Performance Monitoring OpenSBI extension is herein adopted. The added OpenSBI functions are detailed in \autoref{tab:HPMExtension}. It allows to support reading and writing to all the privileged registers defined in version 1.11 of the specification, namely:
\begin{itemize}
    \item Generic Performance counters:  \HPMcounter{mcycle}, \HPMcounter{mtime}, \HPMcounter{minstret}.
    \item Performance counters: \HPMcounter{mhpmcounter\#}.
    \item Event configuration registers: \HPMcounter{mhpmevent\#}.
    \item Lower privilege counter access: \HPMcounter{mcounteren}.
    \item Inhibiting counter increment, \HPMcounter{mcountinhibit}.
\end{itemize}

\begin{table*}[t!]
\centering
\caption{OpenSBI HPM extension function calls definition.}
\label{tab:HPMExtension}
\vspace*{-3mm}
\begin{tabular}{l|l|l|l}
\bf HPM Function                 & \bf Output          & \bf Arguments                & \bf Errors                                                                                                                                     \\ \hline
\rowcolor[HTML]{E0E0E0} 
hpm\_get\_mevent             & event id        & mHPM event id (3 - 31)   & *A: if register not implemented                                                                                      \\
hpm\_set\_mevent             &                 & mHPM event id, event id  & *A: if register not implemented                                                                                      \\
\rowcolor[HTML]{E0E0E0} 
hpm\_get\_{[}m/u{]}counter           & value           & mHPM counter id (0 - 31) & *A: if counter not implemented                                                                                       \\
hpm\_set\_{[}m/u{]}           &                 & mHPM counter id, value   & *A: if counter not implemented                                                                                       \\
\rowcolor[HTML]{E0E0E0} 
hpm\_get\_{[}m/s{]}counteren & 32-bits bitmask &                          & *A: if not implemented                                                                                               \\
hpm\_set\_{[}m/s{]}counteren &                 & 32-bits bitmask          & *A: if not implemented                                                                                               \\
\rowcolor[HTML]{E0E0E0} 
hpm\_get\_mcountinhibit      & 32-bits bitmask &                          & *A: if not implemented                                                                                               \\
hpm\_set\_mcountinhibit      &                 & 32-bits bitmask          & \begin{tabular}[c]{@{}l@{}}*A: if not implemented\\ *B: on trying to inhibit time counter\end{tabular}
\end{tabular}
\begin{tabular}{l}
\rowcolor[HTML]{FFFFFF}
*A: SBI\_ERR\_NOT\_SUPPORTED; *B: SBI\_ERR\_DENIED
\end{tabular}
\vspace*{-2mm}
\end{table*}

Moreover, we also add support for reading and writing directly to the supervisor privilege \HPMcounter{scounteren} register and to the user-level \HPMcounter{hpmcounter} performance counters. While this is not an absolutely necessary feature, considering that the Linux Kernel will have sufficient privilege level, it allows the code to access the counters through an unified interface. This may be deprecated in the future if the performance impact is non-negligible and if there is no actual benefit at software level.

Considering the return structure of the OpenSBI handler for the RISC-V environmental call (\textit{ecall}):
\verbatimfont{\footnotesize}%
\begin{lstlisting}
struct sbi_ret {
    long value;
    long error;
}
\end{lstlisting}
it was determined that each counter/register read could be executed in a single environmental call. Taking into account that the return variable \textit{value} is of type \textit{long}, the variable size will be the same of the implementation scalar registers (i.e., 64-bits in a 64-bit implementation, and 32-bits in a 32-bits implementation). Taking this into account, for a 32-bits system, the process of reading any HPM counter must be unfold in, at least, two calls, reading separately the lower and higher 32-bits portions of \HPMcounter{mhpmcounter\#}. Additional calls may eventually be necessary to compensate for the lower 32-bits counter overflow.
 

The proposed OpenSBI extension will be named HPM, after the RISC-V Hardware Performance Monitor specification, and should be identified by the value 0x48504d (as the direct conversion of "HPM" from ASCII to hexadecimal). Currently, the HPM is experimental, and thus is included in the experimental extension space with the corresponding ID (0x0848504d).

\subsection{Linux Kernel Driver Modifications}
The software-level changes proposed in this work do not impact the majority of the Linux kernel source-code. In particular, they are limited to specific areas, such as the Perf tool code and the Perf related RISC-V kernel portion (arch/riscv/kernel).

As mentioned in the beginning of this section, the current \mbox{RISC-V} Perf kernel implementation gives basic support for the RISC-V HPM specification, being restricted to fixed-event counters, i.e., each event can only be counted from a continuously running, non-stopable, and non-changeable counter. Moreover, the only supported events are the cycle and instret counters, having no process for reading other HPM counters. Hence, it is not compatible with the current RISC-V HPM specification, that allows for event configuration, counter inhibition and to control counter access. However, it does provide a basic structure to work on, which we extend in this work. We also build upon a Request for Comments kernel patch suggested by Zong Li \cite{li_lkml_2020}, that introduced support for the HPM counters, through raw events, and device-tree bindings to support platform-specific hardware events (although such patch was not merged into the kernel).
The introduced support for raw events allows the kernel driver to configure raw, not general, performance monitoring events, providing the necessary interfaces for adding, enabling, disabling and removing events that are related with the HPM counters. In addition, the support for device-tree bindings, allows for each platform to specify its own features, such as:
\begin{itemize}
    \item Width of Base Counters (cycle, time, instret)
    \item Width of Event Counters (\HPMcounter{mhpmevent\#})
    \item Number of Event Counters
    \item Hardware Event Map
    \item Hardware Cache Event Map 
\end{itemize}
The Hardware Event Map is provided as a list of key-value pairs, where each pair matches a hardware event generic to Perf (key) to an implementation raw hardware event (value). An example is to use a key-value pair of \texttt{branch\_misses: 0x05}, where the hardware event 0x05 matches the Perf branch\_misses event. The Hardware Cache Event Map is a similar structure to Hardware Event Map, however it maps events related to cache structures, such as L1 Read Misses or L2 Write Accesses. 

The kernel Perf implementation is responsible for two processes: event sampling (in a general way) and interaction with the CPU Performance Monitoring Unit (through OpenSBI, or directly).
While any event is being sampled, the kernel driver will enable and start an event, proceed to take samples, and then, stop and disable the respective event. The interaction with the CPU PMU is handled at each of the mentioned steps. The introduced modifications include the interaction between the driver and the configuration registers (\HPMcounter{mhpmevent\#}, \HPMcounter{mcounteren}), which are accomplished through OpenSBI, to provide machine-level access. Furthermore, and to decrease performance monitoring overheads, we configure \HPMcounter{mcounteren} to allow for supervisor-level read access, to achieve direct read access from the kernel to the HPM counters.

\begin{figure}[H]
    \centering
    \includegraphics[width=0.44\textwidth]{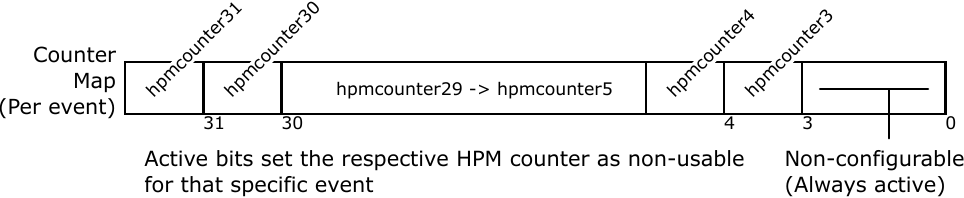}
    \caption{Counter Map for event to HPM counter matching.}
    \label{fig:CounterMap}
\end{figure}

Additionally, we introduced changes to how the events are matched to each counter. While an event could be matched to any counter, where an implementation would provide from 0 to 29 completely generic HPM counters, we alternatively propose that the mapping of each event is constrained to a specific set of counters (providing native support for hardware-friendly implementations where counters and associated events are constrained to specific pipeline stages). To achieve this, any raw event identifier contains two parameters: the event itself and the counter map. The event identifier is a numeric value to be interpreted by the CPU PMU through the \HPMcounter{mhpmevent\#} registers, not constrained to any particular logic (e.g., event classes and sub-classes). In contrast, the counter map is proposed as a 32-bit value, each event has one and only one associated counter map, where each active bit indicates that the corresponding HPM counter is unable to count the event, as depicted in \autoref{fig:CounterMap}. This additional parameter allows any event to be matched to any number and selection of HPM counters.

%

\subsection{Perf Tool Modifications}
The kernel driver is responsible for the actual sampling, the Perf tool acts as the frontend for event counting, providing a user interface for event listing (perf list), performance analysis (perf stat, monitor, record, report), and a set of simple benchmarks (perf bench). The modifications introduced by this work attempt to give support for raw events in a flexible and platform-specific way. In particular, the modifications can be separated in two sets, CPU identification and events mapping.

\begin{figure}[H]
    \centering
    \includegraphics[width=0.45\textwidth]{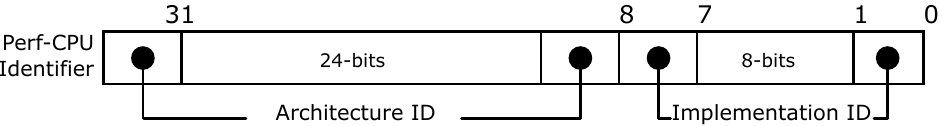}
    \caption{CPU unique identification for Perf events.}
    \label{fig:CPUID}
\end{figure}

To be able to map the set of events available in a specific processor, system or platform, we need to identify which CPU is executing Perf. According to the RISC-V ISA and OpenSBI specifications, each RISC-V implementation has a publicly available architecture ID \cite{risc-v_foundation_open-source_2021}, that is readable through an OpenSBI read of the RISC-V CSR \textit{marchid}. Considering specific implementations can be under the same architecture ID, it is possible to get additional identification of the implemented CPU through another OpenSBI read to the CSR \textit{mimpid}, getting the specific implementation identifier. Taking into account the architecture and implementation identifiers, we consider that an absolute identification can be made by merging together the lower 24-bits of the architecture identifier to the lower 8-bits of the implementation identifier (see \autoref{fig:CPUID}). The choice of value widths can provide up to $2^{24} \approx 17~million$ different architectures and  $2^{8} = 256$ implementations of each architecture, it is expect that these values will not be a future constraint. 

Each CPU Identifier can be mapped to a set of files containing fully described events. This is achieved through a mapping file in CSV format, with the example structure:
\begin{lstlisting}
CPU ID, File Vers. , Events Filename, Events Type
0x300 , 0          , CVA6           , core
0x500 , 0          , SPIKE          , core
0x200 , 0          , BOOM           , core
...
\end{lstlisting}
Where the CPU Identifier is defined according to the mentioned rules, the File Version is currently unused, the Events Filename is set to the name of directory containing the events description, and the Events Type describes the type of events the PMU specifies.
Each directory specified by the Events Filename column can contain multiple files in the Java Script Object Notation (JSON) format. Usually each file describes an event group from one specific category (e.g., pipeline, memory, instructions, etc.), and each of the JSON files will contain one to multiple events, with the following structure:
\begin{lstlisting}
{
    "Public Description": "This is an example event,
    for demonstration purposes.",
    "Brief Description": "This is an example event."
    "Event Code" : "0x11",
    "Counter Mask" : "0xF8FF",
    "Event Name" : "EXAMPLE_EVENT",
}
\end{lstlisting}
For this example, the Event Code 0x11 will be used to configure the \HPMcounter{mhpmevent\#} registers of the available counters selected by the Counter Mask value, where 0xF8FF specifies that the counters 8, 9 and 10 can be used to sample the event.

When monitoring performance, the selected events will be forwarded to the kernel driver, that in turn will handle the event to counter mapping and HPM event configuration. The kernel driver will, in turn, schedule each event or, in alternative, multiplex a set of events in the respective register, allowing for multiple events to be sampled in one workload execution, at the cost of samples accuracy. This process is depicted in \autoref{fig:SystemOverview}.  
    
\subsection{Summary}
The modifications to the kernel, Perf and OpenSBI (enfatised in \autoref{fig:SystemOverview}), allowed us to take advantage of the RISC-V HPM specification. However, the retro-compatibility, and the divergence in the already implemented platforms poses challenges which must be faced during setup and testing of the performance monitoring software. In our current implementation, the hardware event and cache event maps (defined through the device tree bindings) are working as fixed-events, each mapping attaches the respective event to a specific counter. Consequently, this works for defined fixed-event counters, and does not have the flexibility of Perf mapping solution. Therefore, due to the inferior capabilities and increased difficulty of maintaining both solutions, the device-tree configurable events may be removed. Moreover, to better identify which processor features are available, we aim at using OpenSBI to discover registers capabilities and determine, at run-time, which features are implemented. Additionally, and to prevent illegal access to registers, the OpenSBI extension may be modified to prevent accesses that are not available in the platform, returning a descriptive error indicating that the feature is not supported. Through the OpenSBI capabilities and the Perf kernel driver we seek to achieve a good balance between a low performance overhead and powerful performance monitoring capabilities.


\section{Developments and results} \label{sec:Developed}

Initial testing and development was started in QEMU \cite{lupori2018towards,bellard_qemu_2005} through the implementation of version 1.11 of the privileged HPM specification. While QEMU is a flexible, fast and powerful platform for software development, mainly when there is lack of widely available hardware, it is difficult to emulate specific hardware. In particular, while there were no major difficulties in providing read and write access to the emulated configuration registers and performance counters, counting actual performance values proved unfeasible in practice.

\begin{table}[t]
\centering
\caption{Available fixed performance events for the CVA6 platform.}
\label{tab:CVA6-Counters}
\vspace*{-2mm}
\begin{tabular}{l|l}
Event                   & Counter       \\ \hline
Cycles                  & mcycle        \\
Instructions Retired    & minstret      \\
ICache Misses           & mhpmcounter3  \\
DCache Misses           & mhpmcounter4  \\
ITLB Misses             & mhpmcounter5  \\
DTLB Misses             & mhpmcounter6  \\
Loads                   & mhpmcounter7  \\
Stores                  & mhpmcounter8  \\
Taken Exceptions        & mhpmcounter9  \\
Exceptions Returned     & mhpmcounter10 \\
Branches and Jumps      & mhpmcounter11 \\
Calls                   & mhpmcounter12 \\
Returns                 & mhpmcounter13 \\
Mispredicted Branches   & mhpmcounter14 \\
Scoreboard Full         & mhpmcounter15 \\
Instruction Fetch Empty & mhpmcounter16
\end{tabular}
\vspace*{-3mm}
\end{table}

The first attempt of emulating performance counters was to increase each counter by a randomized value on each counter read. While the strategy was simple, there was no validity to the obtained results. A temporary alternative solution was to pause the emulated system, execute the gem5 simulator \cite{lowe-power_gem5_2020} with the corresponding application, and use the filesystem to retrieve gem5 statistics onto QEMU. Evidently, pausing QEMU execution during a performance counter evaluation was not feasible, and although the results could have some validity, the counter sampling sequence was not realistic, as only one of the samples would actually increment. Another solution could be to execute an actual performance monitoring of the host system through Perf, although intuitively this would result in the same problems of the prior solution. Considering these limitations, QEMU was used to develop the majority of software and to perform initial testing.

\begin{table*}[t]
\centering
\caption{Computed metrics from performance monitoring of the CoreMark benchmark.}
\label{tab:Metrics}
\begin{tabular}{ll|l}
\multicolumn{2}{c}{Metric}                   & \multicolumn{1}{c}{Events}                \\ \hline
Branch MissRate                     & 18.14\% & Mispredictions / Branches, Calls, Returns \\
L1D MissRate                        & 0.95\%  & L1D Misses / Loads, Stores                \\
L1I MissRate                        & 0.58\%  & L1I Misses / Instructions                 \\
ScoreBoard Full (cycles)            & 0.38\%  & ScoreBoard Full Cycles / Cycles           \\
Instruction Fetch Empty (cycles)    & 10.12\% & IF Empty Cycles / Cycles                  \\
Instructions Per Cycle              & 0.6195 & Instructions / Cycles                     \\
Translation MissRate (Data)         & 0.00\%  & Data TLB Misses / Loads, Stores           \\
Translation MissRate (Instructions) & 0.47\%  & Instructions TLB Misses / Instructions   
\end{tabular}
\end{table*}

To evaluate the solution in a real RISC-V platform, we set with the CVA6 \cite{zaruba_cost_2019} core (previously named Ariane), executing on an FPGA at 100~MHz, running Linux 5.7.0 with BusyBox 1.31.1. The choice of system was due to availability and we expect to widen the testing platforms in the future. Currently, our CVA6 implementation only supports the fixed-event counters detailed in \autoref{tab:CVA6-Counters}.

With our extensions and modifications, it is now possible to list these events when running \textit{perf list} on top of the CVA6 implementation. Outputting the following list of available events (shortened for better representation):
\begin{lstlisting}
  branch-instructions OR branches   [Hardware event]
  branch-misses                     [Hardware event]
  cache-misses                      [Hardware event]
  ...
  alignment-faults                  [Software event]
  ...
  iTLB-load-misses            [Hardware cache event]

branch:
  ariane_branch_jump
       [Branches/jumps count]
  ...
  ariane_ret
       [Returns count]

cache:
  ariane_dtlb_miss
       [Data TLB miss]
  ...
  ariane_store
       [Data loads]

pipeline:
  ariane_exception
       [Exceptions count]
  ...
  riscv_cycles
       [CPU cycles]
\end{lstlisting}

The CoreMark \cite{gal-on_exploring_2012} benchmark was used to further test through actual performance monitoring. Version 1.0 of CoreMark was executed in the CVA6 core and monitored under (perf stat), achieving a performance of 174.59 points at 100~MHz. The Perf \textit{stat} command reported the following monitored events during execution:  
\begin{lstlisting}
     Performance counter stats for '/bin/coremark':
         236011286      ariane_branch_jump
           5312578      ariane_call
          44038701      ariane_mis_predict
           1406812      ariane_ret
              1118      ariane_dtlb_miss
           6869722      ariane_itlb_miss
           2786559      ariane_l1_dcache_miss
           8443755      ariane_l1_icache_miss
         229104327      ariane_load
          64628214      ariane_store
             22486      ariane_exception
             22486      ariane_exception_ret
         239773306      ariane_if_empty
           9094173      ariane_sb_full
        2368685119      riscv_cycles
        1467339227      riscv_instret

      23.779291520 seconds time elapsed

      23.578690000 seconds user
       0.139518000 seconds sys
\end{lstlisting}
By adding metrics support in Perf, the values in \autoref{tab:Metrics} can be automatically computed after monitoring an application performance. Through multiple executions of the CoreMark benchmark, with and without Perf monitoring, the performance penalty of using Perf was determined as 0.283\%, for this benchmark.
\section{Conclusions and Next Steps} \label{sec:Conclusions}
In this paper a RISC-V compatible performance monitoring solution is proposed, allowing for developers to do platform-specific code optimization for RISC-V processors. While there was already initial support for RISC-V performance monitoring through the Performance analysis tools for Linux (Perf), it was not comprehensive enough and supported only cycles and instructions counting. Although that version of Perf did not support the latest ISA specification, we introduce support through simple modifications to the Perf Kernel driver and to the Perf application, alongside with a new extension to OpenSBI that enables interaction with higher privilege configuration registers.

Despite Perf utilization being declining over time, mainly due to the takeover of Intel and ARM proprietary platforms, we consider this as an opportunity for Perf. Considering the open-source nature of RISC-V, it is only natural to rely on open-source tools. At this concern, Perf is a powerful, extendable and comprehensive performance monitoring tool for RISC-V, provides the natural interface for higher level libraries (e.g., PAPI).

While this proposal tackles the interface between events and the RISC-V hardware performance monitoring facility, there are other open-source initiatives, through Kernel patches, that seek to improve Perf with alternative or complimentary objectives (e.g., supporting firmware level events). Accounting that this paper represents an early work, the presented proposal may be improved by merging with upcoming opportunities or diverging ideas. Furthermore, we seek to further test the proposed performance monitoring approach and to apply it in ASIC RISC-V hardware, such as the Hi-Five Boards and the upcoming BeagleBones BeagleV board.


\begin{acks}
This work was partially supported by national funds through Fundação para a Ciência e a Tecnologia (FCT) under projects\\
UIDB/50021/2020, and by funds from the European Union Horizon 2020 Research and Innovation programme under grant agreement No. 826647, European Processor Initiative EPI.
We acknowledge the Institute of Computer Science (ICS) team at the Foundation for Research and Technology - Hellas (FORTH), in particular to the Computer Architecture and VLSI Systems (CARV) personel: Nick Kossifidis, Georgios Ieronymakis, Nikolaus Dimou, and Vassilis Papaefstathiou, for all the heavily appreciated support, tools and guidance.
\end{acks}

\balance
\bibliographystyle{ACM-Reference-Format}
\bibliography{refs}


\begin{thebibliography}{30}


\ifx \showCODEN    \undefined \def \showCODEN     #1{\unskip}     \fi
\ifx \showDOI      \undefined \def \showDOI       #1{#1}\fi
\ifx \showISBNx    \undefined \def \showISBNx     #1{\unskip}     \fi
\ifx \showISBNxiii \undefined \def \showISBNxiii  #1{\unskip}     \fi
\ifx \showISSN     \undefined \def \showISSN      #1{\unskip}     \fi
\ifx \showLCCN     \undefined \def \showLCCN      #1{\unskip}     \fi
\ifx \shownote     \undefined \def \shownote      #1{#1}          \fi
\ifx \showarticletitle \undefined \def \showarticletitle #1{#1}   \fi
\ifx \showURL      \undefined \def \showURL       {\relax}        \fi
\providecommand\bibfield[2]{#2}
\providecommand\bibinfo[2]{#2}
\providecommand\natexlab[1]{#1}
\providecommand\showeprint[2][]{arXiv:#2}

\bibitem[\protect\citeauthoryear{Balkind, Lim, Gao, Tu, Wentzlaff, Schaffner,
  Zaruba, and Benini}{Balkind et~al\mbox{.}}{2019}]%
        {balkind_openpitonariane_2019}
\bibfield{author}{\bibinfo{person}{Jonathan Balkind}, \bibinfo{person}{Katie
  Lim}, \bibinfo{person}{Fei Gao}, \bibinfo{person}{Jinzheng Tu},
  \bibinfo{person}{David Wentzlaff}, \bibinfo{person}{Michael Schaffner},
  \bibinfo{person}{Florian Zaruba}, {and} \bibinfo{person}{Luca Benini}.}
  \bibinfo{year}{2019}\natexlab{}.
\newblock \showarticletitle{{OpenPiton}+{Ariane}: {The} {First}
  {Open}-{Source}, {SMP} {Linux}-booting {RISC}-{V} {System} {Scaling} {From}
  {One} to {Many} {Cores}}. In \bibinfo{booktitle}{\emph{Workshop on {Computer}
  {Architecture} {Research} with {RISC}-{V} ({CARRV})}}. \bibinfo{pages}{1--6}.
\newblock


\bibitem[\protect\citeauthoryear{Bellard}{Bellard}{2005}]%
        {bellard_qemu_2005}
\bibfield{author}{\bibinfo{person}{Fabrice Bellard}.}
  \bibinfo{year}{2005}\natexlab{}.
\newblock \showarticletitle{{QEMU}, a fast and portable dynamic translator.}.
  In \bibinfo{booktitle}{\emph{{USENIX} annual technical conference, {FREENIX}
  {Track}}}, Vol.~\bibinfo{volume}{41}. \bibinfo{publisher}{Califor-nia, USA},
  \bibinfo{pages}{46}.
\newblock


\bibitem[\protect\citeauthoryear{Celio, Chiu, Nikolic, Patterson, and
  Asanovic}{Celio et~al\mbox{.}}{2017}]%
        {celio_boom_2017}
\bibfield{author}{\bibinfo{person}{Christopher Celio}, \bibinfo{person}{Pi-Feng
  Chiu}, \bibinfo{person}{Borivoje Nikolic}, \bibinfo{person}{David Patterson},
  {and} \bibinfo{person}{Krste Asanovic}.} \bibinfo{year}{2017}\natexlab{}.
\newblock \bibinfo{booktitle}{\emph{{BOOM} v2: an open-source out-of-order
  {RISC}-{V} core}}.
\newblock \bibinfo{type}{{T}echnical {R}eport} UCB/EECS-2017-157.
  \bibinfo{institution}{University of California at Berkeley}.
  \bibinfo{pages}{8} pages.
\newblock


\bibitem[\protect\citeauthoryear{Chen, Xiang, Liu, Shang, Guo, Liu, Lu, Hao,
  Luo, Chen, Li, Pu, Meng, Yan, Xie, and Qi}{Chen et~al\mbox{.}}{2020}]%
        {chen_xuantie-910_2020}
\bibfield{author}{\bibinfo{person}{C. Chen}, \bibinfo{person}{X. Xiang},
  \bibinfo{person}{C. Liu}, \bibinfo{person}{Y. Shang}, \bibinfo{person}{R.
  Guo}, \bibinfo{person}{D. Liu}, \bibinfo{person}{Y. Lu}, \bibinfo{person}{Z.
  Hao}, \bibinfo{person}{J. Luo}, \bibinfo{person}{Z. Chen},
  \bibinfo{person}{C. Li}, \bibinfo{person}{Y. Pu}, \bibinfo{person}{J. Meng},
  \bibinfo{person}{X. Yan}, \bibinfo{person}{Y. Xie}, {and} \bibinfo{person}{X.
  Qi}.} \bibinfo{year}{2020}\natexlab{}.
\newblock \showarticletitle{Xuantie-910: {A} {Commercial} {Multi}-{Core}
  12-{Stage} {Pipeline} {Out}-of-{Order} 64-bit {High} {Performance} {RISC}-{V}
  {Processor} with {Vector} {Extension} : {Industrial} {Product}}. In
  \bibinfo{booktitle}{\emph{2020 {ACM}/{IEEE} 47th {Annual} {International}
  {Symposium} on {Computer} {Architecture} ({ISCA})}}. \bibinfo{pages}{52--64}.
\newblock
\urldef\tempurl%
\url{https://doi.org/10.1109/ISCA45697.2020.00016}
\showDOI{\tempurl}


\bibitem[\protect\citeauthoryear{De~Melo}{De~Melo}{2010}]%
        {de_melo_new_2010}
\bibfield{author}{\bibinfo{person}{Arnaldo~Carvalho De~Melo}.}
  \bibinfo{year}{2010}\natexlab{}.
\newblock \bibinfo{title}{The new linux’perf’tools}.
\newblock
\newblock
\urldef\tempurl%
\url{http://www.linux-kongress.org/2010/slides/lk2010-perf-acme.pdf}
\showURL{%
\tempurl}


\bibitem[\protect\citeauthoryear{Dongarra, London, Moore, Mucci, and
  Terpstra}{Dongarra et~al\mbox{.}}{2001}]%
        {dongarra_using_2001}
\bibfield{author}{\bibinfo{person}{Jack Dongarra}, \bibinfo{person}{Kevin
  London}, \bibinfo{person}{Shirley Moore}, \bibinfo{person}{Phil Mucci}, {and}
  \bibinfo{person}{Dan Terpstra}.} \bibinfo{year}{2001}\natexlab{}.
\newblock \showarticletitle{Using {PAPI} for {Hardware} {Performance}
  {Monitoring} on {Linux} {Systems}}. In \bibinfo{booktitle}{\emph{In
  {Conference} on {Linux} {Clusters}: {The} {HPC} {Revolution}, {Linux}
  {Clusters} {Institute}}}.
\newblock


\bibitem[\protect\citeauthoryear{Gal-On and Levy}{Gal-On and Levy}{2012}]%
        {gal-on_exploring_2012}
\bibfield{author}{\bibinfo{person}{Shay Gal-On} {and} \bibinfo{person}{Markus
  Levy}.} \bibinfo{year}{2012}\natexlab{}.
\newblock \showarticletitle{Exploring coremark a benchmark maximizing
  simplicity and efficacy}.
\newblock \bibinfo{journal}{\emph{The Embedded Microprocessor Benchmark
  Consortium}} (\bibinfo{year}{2012}).
\newblock


\bibitem[\protect\citeauthoryear{{Intel}}{{Intel}}{2016}]%
        {intel_intel_2016}
\bibfield{author}{\bibinfo{person}{{Intel}}.} \bibinfo{year}{2016}\natexlab{}.
\newblock \bibinfo{booktitle}{\emph{Intel® 64 and {IA}-32 {Architectures}
  {Developer}'s {Manual}: {Vol}. {3B}}}.
\newblock \bibinfo{type}{{T}echnical {R}eport}.
\newblock
\urldef\tempurl%
\url{https://www.intel.com/content/www/us/en/architecture-and-technology/64-ia-32-architectures-software-developer-vol-3b-part-2-manual.html}
\showURL{%
\tempurl}


\bibitem[\protect\citeauthoryear{Jagode, Danalis, Anzt, and Dongarra}{Jagode
  et~al\mbox{.}}{2019}]%
        {jagode_papi_2019}
\bibfield{author}{\bibinfo{person}{Heike Jagode}, \bibinfo{person}{Anthony
  Danalis}, \bibinfo{person}{Hartwig Anzt}, {and} \bibinfo{person}{Jack
  Dongarra}.} \bibinfo{year}{2019}\natexlab{}.
\newblock \showarticletitle{{PAPI} software-defined events for in-depth
  performance analysis}.
\newblock \bibinfo{journal}{\emph{The International Journal of High Performance
  Computing Applications}} \bibinfo{volume}{33}, \bibinfo{number}{6}
  (\bibinfo{date}{Nov.} \bibinfo{year}{2019}), \bibinfo{pages}{1113--1127}.
\newblock
\showISSN{1094-3420}
\urldef\tempurl%
\url{https://doi.org/10.1177/1094342019846287}
\showDOI{\tempurl}
\newblock
\shownote{Publisher: SAGE Publications Ltd STM.}


\bibitem[\protect\citeauthoryear{Kao and {The kernel development
  community}}{Kao and {The kernel development community}}{2018}]%
        {kao_supporting_2018}
\bibfield{author}{\bibinfo{person}{Alan Kao} {and} \bibinfo{person}{{The kernel
  development community}}.} \bibinfo{year}{2018}\natexlab{}.
\newblock \bibinfo{title}{Supporting {PMUs} on {RISC}-{V} platforms — {The}
  {Linux} {Kernel} documentation}.
\newblock
\newblock
\urldef\tempurl%
\url{https://www.kernel.org/doc/html/latest/riscv/pmu.html}
\showURL{%
\tempurl}


\bibitem[\protect\citeauthoryear{Kleen and Strong}{Kleen and Strong}{2015}]%
        {kleen_intel_2015}
\bibfield{author}{\bibinfo{person}{Andi Kleen} {and} \bibinfo{person}{Beeman
  Strong}.} \bibinfo{year}{2015}\natexlab{}.
\newblock \showarticletitle{Intel ® {Processor} {Trace} on {Linux}}.
\newblock \bibinfo{journal}{\emph{Tracing Summit}}  \bibinfo{volume}{2015}
  (\bibinfo{year}{2015}).
\newblock
\urldef\tempurl%
\url{https://citeseerx.ist.psu.edu/viewdoc/download?doi=10.1.1.735.3516&rep=rep1&type=pdf}
\showURL{%
\tempurl}


\bibitem[\protect\citeauthoryear{Lee, Lee, Heo, Hwang, and Paek}{Lee
  et~al\mbox{.}}{2017}]%
        {lee_using_2017}
\bibfield{author}{\bibinfo{person}{Yongje Lee}, \bibinfo{person}{Jinyong Lee},
  \bibinfo{person}{Ingoo Heo}, \bibinfo{person}{Dongil Hwang}, {and}
  \bibinfo{person}{Yunheung Paek}.} \bibinfo{year}{2017}\natexlab{}.
\newblock \showarticletitle{Using {CoreSight} {PTM} to {Integrate} {CRA}
  {Monitoring} {IPs} in an {ARM}-{Based} {SoC}}.
\newblock \bibinfo{journal}{\emph{ACM Transactions on Design Automation of
  Electronic Systems}} \bibinfo{volume}{22}, \bibinfo{number}{3}
  (\bibinfo{date}{April} \bibinfo{year}{2017}), \bibinfo{pages}{52:1--52:25}.
\newblock
\showISSN{1084-4309}
\urldef\tempurl%
\url{https://doi.org/10.1145/3035965}
\showDOI{\tempurl}


\bibitem[\protect\citeauthoryear{Li}{Li}{2020}]%
        {li_lkml_2020}
\bibfield{author}{\bibinfo{person}{Zong Li}.} \bibinfo{year}{2020}\natexlab{}.
\newblock \bibinfo{title}{{LKML}: {Zong} {Li}: [{RFC} {PATCH} 0/6] {Support}
  raw event and {DT} for perf on {RISC}-{V}}.
\newblock
\newblock
\urldef\tempurl%
\url{https://lkml.org/lkml/2020/6/28/374}
\showURL{%
\tempurl}


\bibitem[\protect\citeauthoryear{London, Moore, Mucci, Seymour, and
  Luczak}{London et~al\mbox{.}}{2001}]%
        {london_papi_2001}
\bibfield{author}{\bibinfo{person}{Kevin London}, \bibinfo{person}{Shirley
  Moore}, \bibinfo{person}{Philip Mucci}, \bibinfo{person}{Keith Seymour},
  {and} \bibinfo{person}{Richard Luczak}.} \bibinfo{year}{2001}\natexlab{}.
\newblock \showarticletitle{The {PAPI} {Cross}-{Platform} {Interface} to
  {Hardware} {Performance} {Counters}}. In \bibinfo{booktitle}{\emph{Department
  of {Defense} {Users}’ {Group} {Conference} {Proceedings}.}}
  \bibinfo{pages}{18--21}.
\newblock


\bibitem[\protect\citeauthoryear{Lowe-Power, Ahmad, Akram, Alian, Amslinger,
  Andreozzi, Armejach, Asmussen, Beckmann, Bharadwaj, and {others}}{Lowe-Power
  et~al\mbox{.}}{2020}]%
        {lowe-power_gem5_2020}
\bibfield{author}{\bibinfo{person}{Jason Lowe-Power},
  \bibinfo{person}{Abdul~Mutaal Ahmad}, \bibinfo{person}{Ayaz Akram},
  \bibinfo{person}{Mohammad Alian}, \bibinfo{person}{Rico Amslinger},
  \bibinfo{person}{Matteo Andreozzi}, \bibinfo{person}{Adrià Armejach},
  \bibinfo{person}{Nils Asmussen}, \bibinfo{person}{Brad Beckmann},
  \bibinfo{person}{Srikant Bharadwaj}, {and} \bibinfo{person}{{others}}.}
  \bibinfo{year}{2020}\natexlab{}.
\newblock \showarticletitle{The gem5 simulator: {Version} 20.0+}.
\newblock \bibinfo{journal}{\emph{arXiv preprint arXiv:2007.03152}}
  (\bibinfo{year}{2020}).
\newblock


\bibitem[\protect\citeauthoryear{Lupori, Rosario, and Borin}{Lupori
  et~al\mbox{.}}{2018}]%
        {lupori2018towards}
\bibfield{author}{\bibinfo{person}{Leandro Lupori}, \bibinfo{person}{Vanderson
  Rosario}, {and} \bibinfo{person}{Edson Borin}.}
  \bibinfo{year}{2018}\natexlab{}.
\newblock \showarticletitle{Towards a high-performance {RISC}-{V} emulator}. In
  \bibinfo{booktitle}{\emph{2018 symposium on high performance computing
  systems ({WSCAD})}}. \bibinfo{pages}{213--220}.
\newblock
\newblock
\shownote{tex.organization: IEEE.}


\bibitem[\protect\citeauthoryear{Marques, Ilic, Matveev, and Sousa}{Marques
  et~al\mbox{.}}{2020}]%
        {marques2020application}
\bibfield{author}{\bibinfo{person}{Diogo Marques}, \bibinfo{person}{Aleksandar
  Ilic}, \bibinfo{person}{Zakhar~A Matveev}, {and} \bibinfo{person}{Leonel
  Sousa}.} \bibinfo{year}{2020}\natexlab{}.
\newblock \showarticletitle{Application-driven cache-aware roofline model}.
\newblock \bibinfo{journal}{\emph{Future Generation Computer Systems}}
  \bibinfo{volume}{107} (\bibinfo{year}{2020}), \bibinfo{pages}{257--273}.
\newblock
\newblock
\shownote{Publisher: Elsevier.}


\bibitem[\protect\citeauthoryear{Matthews and Shannon}{Matthews and
  Shannon}{2017}]%
        {matthews_taiga_2017}
\bibfield{author}{\bibinfo{person}{Eric Matthews} {and} \bibinfo{person}{Lesley
  Shannon}.} \bibinfo{year}{2017}\natexlab{}.
\newblock \showarticletitle{{TAIGA}: {A} new {RISC}-{V} soft-processor
  framework enabling high performance {CPU} architectural features}. In
  \bibinfo{booktitle}{\emph{2017 27th {International} {Conference} on {Field}
  {Programmable} {Logic} and {Applications} ({FPL})}}. \bibinfo{pages}{1--4}.
\newblock
\urldef\tempurl%
\url{https://doi.org/10.23919/FPL.2017.8056766}
\showDOI{\tempurl}
\newblock
\shownote{ISSN: 1946-1488.}


\bibitem[\protect\citeauthoryear{{OProfile}}{{OProfile}}{2021}]%
        {oprofile_oprofile_2021}
\bibfield{author}{\bibinfo{person}{{OProfile}}.}
  \bibinfo{year}{2021}\natexlab{}.
\newblock \bibinfo{title}{{OProfile} - {A} {System} {Profiler} for {Linux}}.
\newblock
\newblock
\urldef\tempurl%
\url{https://oprofile.sourceforge.io/about/}
\showURL{%
\tempurl}


\bibitem[\protect\citeauthoryear{{RISC-V Foundation}}{{RISC-V
  Foundation}}{2021}]%
        {risc-v_foundation_open-source_2021}
\bibfield{author}{\bibinfo{person}{{RISC-V Foundation}}.}
  \bibinfo{year}{2021}\natexlab{}.
\newblock \bibinfo{title}{Open-{Source} {RISC}-{V} {Architecture} {IDs}}.
\newblock
\newblock
\urldef\tempurl%
\url{https://github.com/riscv/riscv-isa-manual/blob/master/marchid.md}
\showURL{%
\tempurl}


\bibitem[\protect\citeauthoryear{{SiFive}}{{SiFive}}{2018}]%
        {sifive_sifive_2018}
\bibfield{author}{\bibinfo{person}{{SiFive}}.} \bibinfo{year}{2018}\natexlab{}.
\newblock \bibinfo{title}{The {SiFive} {HiFive} {Unleashed} {RISC}-{V}
  {Board}}.
\newblock
\newblock
\urldef\tempurl%
\url{https://www.sifive.com/boards/hifive-unleashed}
\showURL{%
\tempurl}


\bibitem[\protect\citeauthoryear{Su, Kuo, Lee, Huang, Jian, Chien, Guo, and
  Chen}{Su et~al\mbox{.}}{2011}]%
        {su_multi-core_2011}
\bibfield{author}{\bibinfo{person}{Alan~P Su}, \bibinfo{person}{Jiff Kuo},
  \bibinfo{person}{Kuen-Jong Lee}, \bibinfo{person}{Ing-Jer Huang},
  \bibinfo{person}{Guo-An Jian}, \bibinfo{person}{Cheng-An Chien},
  \bibinfo{person}{Jiun-In Guo}, {and} \bibinfo{person}{Chien-Hung Chen}.}
  \bibinfo{year}{2011}\natexlab{}.
\newblock \showarticletitle{Multi-core software/hardware co-debug platform with
  {ARM} {CoreSight}™, on-chip test architecture and {AXI}/{AHB} bus monitor}.
  In \bibinfo{booktitle}{\emph{Proceedings of 2011 {International} {Symposium}
  on {VLSI} {Design}, {Automation} and {Test}}}. \bibinfo{pages}{1--6}.
\newblock
\urldef\tempurl%
\url{https://doi.org/10.1109/VDAT.2011.5783594}
\showDOI{\tempurl}


\bibitem[\protect\citeauthoryear{Waterman and Asanović}{Waterman and
  Asanović}{2019}]%
        {watermanRISCVInstructionSet2019a}
\bibfield{author}{\bibinfo{person}{Andrew Waterman} {and}
  \bibinfo{person}{Krste Asanović}.} \bibinfo{year}{2019}\natexlab{}.
\newblock \showarticletitle{The {RISC}-{V} {Instruction} {Set} {Manual},
  {Volume} {II}: {Privileged} {Architecture}, {Version}
  20190608-{Priv}-{MSU}-{Ratified}}.
\newblock \bibinfo{journal}{\emph{RISC-V Foundation}} (\bibinfo{date}{June}
  \bibinfo{year}{2019}).
\newblock


\bibitem[\protect\citeauthoryear{Waterman, Lee, Avizienis, Patterson, and
  Asanović}{Waterman et~al\mbox{.}}{2016a}]%
        {waterman_risc-v_2016}
\bibfield{author}{\bibinfo{person}{Andrew Waterman}, \bibinfo{person}{Yunsup
  Lee}, \bibinfo{person}{Rimas Avizienis}, \bibinfo{person}{David Patterson},
  {and} \bibinfo{person}{Krste Asanović}.} \bibinfo{year}{2016}\natexlab{a}.
\newblock \showarticletitle{The {RISC}-{V} {Instruction} {Set} {Manual},
  {Volume} {II}: {Privileged} {Architecture}, {Document} {Version} 1.9}.
\newblock \bibinfo{journal}{\emph{EECS Department, UC Berkeley, Tech. Rep.
  UCB/EECS-2016-129}}  \bibinfo{volume}{129} (\bibinfo{year}{2016}).
\newblock


\bibitem[\protect\citeauthoryear{Waterman, Lee, Avizienis, Patterson, and
  Asanovic}{Waterman et~al\mbox{.}}{2015}]%
        {waterman_risc-v_2015}
\bibfield{author}{\bibinfo{person}{Andrew Waterman}, \bibinfo{person}{Yunsup
  Lee}, \bibinfo{person}{Rimas Avizienis}, \bibinfo{person}{David~A Patterson},
  {and} \bibinfo{person}{Krste Asanovic}.} \bibinfo{year}{2015}\natexlab{}.
\newblock \showarticletitle{The {RISC}-{V} {Instruction} {Set} {Manual},
  {Volume} {II}: {Privileged} {Architecture}, {Document} {Version} 1.7}.
\newblock \bibinfo{journal}{\emph{EECS Department, UC Berkeley, Tech. Rep.
  UCB/EECS-2015-49}}  \bibinfo{volume}{49} (\bibinfo{year}{2015}).
\newblock


\bibitem[\protect\citeauthoryear{Waterman, Lee, Patterson, and
  Asanovic}{Waterman et~al\mbox{.}}{2016b}]%
        {waterman_risc-v_2016-1}
\bibfield{author}{\bibinfo{person}{Andrew Waterman}, \bibinfo{person}{Yunsup
  Lee}, \bibinfo{person}{David~A Patterson}, {and} \bibinfo{person}{Krste
  Asanovic}.} \bibinfo{year}{2016}\natexlab{b}.
\newblock \showarticletitle{The {RISC}-{V} {Instruction} {Set} {Manual},
  {Volume} {I}: {User}-{Level} {ISA} {Version} 2.1}.
\newblock \bibinfo{journal}{\emph{EECS Department, UC Berkeley, Tech. Rep.
  UCB/EECS-2016-118}}  \bibinfo{volume}{118} (\bibinfo{year}{2016}).
\newblock


\bibitem[\protect\citeauthoryear{Weaver}{Weaver}{2013}]%
        {weaver_linux_2013}
\bibfield{author}{\bibinfo{person}{Vincent~M Weaver}.}
  \bibinfo{year}{2013}\natexlab{}.
\newblock \showarticletitle{Linux perf\_event features and overhead}. In
  \bibinfo{booktitle}{\emph{The 2nd {International} {Workshop} on {Performance}
  {Analysis} of {Workload} {Optimized} {Systems}, {FastPath}}},
  Vol.~\bibinfo{volume}{13}. \bibinfo{pages}{5}.
\newblock


\bibitem[\protect\citeauthoryear{Zaruba and Benini}{Zaruba and Benini}{2019}]%
        {zaruba_cost_2019}
\bibfield{author}{\bibinfo{person}{Florian Zaruba} {and} \bibinfo{person}{Luca
  Benini}.} \bibinfo{year}{2019}\natexlab{}.
\newblock \showarticletitle{The {Cost} of {Application}-{Class} {Processing}:
  {Energy} and {Performance} {Analysis} of a {Linux}-{Ready} 1.7-{GHz} 64-{Bit}
  {RISC}-{V} {Core} in 22-nm {FDSOI} {Technology}}.
\newblock \bibinfo{journal}{\emph{IEEE Transactions on Very Large Scale
  Integration (VLSI) Systems}} \bibinfo{volume}{27}, \bibinfo{number}{11}
  (\bibinfo{date}{Nov.} \bibinfo{year}{2019}), \bibinfo{pages}{2629--2640}.
\newblock
\showISSN{1557-9999}
\urldef\tempurl%
\url{https://doi.org/10.1109/TVLSI.2019.2926114}
\showDOI{\tempurl}
\newblock
\shownote{Conference Name: IEEE Transactions on Very Large Scale Integration
  (VLSI) Systems.}


\bibitem[\protect\citeauthoryear{Zeinolabedin, Partzsch, and Mayr}{Zeinolabedin
  et~al\mbox{.}}{2021}]%
        {zeinolabedin_real-time_2021}
\bibfield{author}{\bibinfo{person}{Seyed Mohammad~Ali Zeinolabedin},
  \bibinfo{person}{Johannes Partzsch}, {and} \bibinfo{person}{Christian Mayr}.}
  \bibinfo{year}{2021}\natexlab{}.
\newblock \showarticletitle{Real-time {Hardware} {Implementation} of {ARM}
  {CoreSight} {Trace} {Decoder}}.
\newblock \bibinfo{journal}{\emph{IEEE Design Test}} \bibinfo{volume}{38},
  \bibinfo{number}{1} (\bibinfo{date}{Feb.} \bibinfo{year}{2021}),
  \bibinfo{pages}{69--77}.
\newblock
\showISSN{2168-2364}
\urldef\tempurl%
\url{https://doi.org/10.1109/MDAT.2020.3002145}
\showDOI{\tempurl}
\newblock
\shownote{Conference Name: IEEE Design Test.}


\bibitem[\protect\citeauthoryear{Zhao, Korpan, Gonzalez, and Asanovic}{Zhao
  et~al\mbox{.}}{2020}]%
        {zhao_sonicboom_2020}
\bibfield{author}{\bibinfo{person}{Jerry Zhao}, \bibinfo{person}{Ben Korpan},
  \bibinfo{person}{Abraham Gonzalez}, {and} \bibinfo{person}{Krste Asanovic}.}
  \bibinfo{year}{2020}\natexlab{}.
\newblock \showarticletitle{{SonicBOOM}: {The} 3rd {Generation} {Berkeley}
  {Out}-of-{Order} {Machine}}.
\newblock \bibinfo{journal}{\emph{Workshop on Computer Architecture Research
  with RISC-V (CARRV)}} (\bibinfo{year}{2020}), \bibinfo{pages}{7}.
\newblock


\end{thebibliography}

\end{document}